\documentclass[%
 reprint,
 amsmath,amssymb,
 aps,
]{revtex4-2}

\usepackage{xcolor}
\usepackage{array}
\usepackage{comment}
\usepackage{graphicx}
\usepackage{dcolumn}
\usepackage{bm}
\usepackage{hyperref}
\hypersetup{
    colorlinks=true,
    linkcolor=blue,
    urlcolor=blue,
    citecolor=blue
}
\usepackage{enumitem}
\usepackage{cleveref}
\usepackage{booktabs}
\usepackage{multirow}
\usepackage{ragged2e}
\usepackage{makecell}

\newcommand{\wcell}[2]{\parbox[t]{#1}{\RaggedRight #2}}

\renewcommand{\arraystretch}{1.1}

\begin{document}

\title{The Grand Challenge of Quantum Applications}

\author{Ryan Babbush}
\email{babbush@google.com}
\affiliation{Google Quantum AI, Santa Barbara, CA 93111, United States}
\author{Robbie King}
\email{robbieking@google.com}
\affiliation{Google Quantum AI, Santa Barbara, CA 93111, United States}
\author{Sergio Boixo}
\affiliation{Google Quantum AI, Santa Barbara, CA 93111, United States}
\author{William Huggins}
\affiliation{Google Quantum AI, Santa Barbara, CA 93111, United States}
\author{Tanuj Khattar}
\affiliation{Google Quantum AI, Santa Barbara, CA 93111, United States}
\author{Guang Hao Low}
\affiliation{Google Quantum AI, Santa Barbara, CA 93111, United States}
\author{Jarrod R.~McClean}
\affiliation{Google Quantum AI, Santa Barbara, CA 93111, United States}
\author{Thomas O'Brien}
\affiliation{Google Quantum AI, Santa Barbara, CA 93111, United States}
\author{Nicholas C.~Rubin}
\affiliation{Google Quantum AI, Santa Barbara, CA 93111, United States}

\date{\today}

\begin{abstract}
This perspective outlines promising pathways and critical obstacles on the road to developing useful quantum computing applications, drawing on insights from the Google Quantum AI team. We propose a five-stage framework for this process, spanning from theoretical explorations of quantum advantage to the practicalities of compilation and resource estimation. For each stage, we discuss key trends, milestones, and inherent scientific and sociological impediments. We argue that two central stages—identifying concrete problem instances expected to exhibit quantum advantage, and connecting such problems to real-world use cases—represent essential and currently under-resourced challenges. Throughout, we touch upon related topics, including the promise of generative artificial intelligence for aspects of this research, criteria for compelling demonstrations of quantum advantage, and the future of compilation as we enter the era of early fault-tolerant quantum computing.
\end{abstract}

\maketitle

\section*{Introduction}

Quantum computing hardware is advancing at a remarkable rate. The theoretical basis of quantum error-correction is solid, and several platforms are now below the error-correction threshold \cite{Bluvstein2024Logical, Sivak2023Real, Suppressing2023Quantum}. Experimentalists believe today’s technology can scale to at least hundreds of logical qubits, and while significant challenges remain, there is growing confidence that there are no fundamentally insurmountable obstacles on the path to a large fault-tolerant quantum computer.

Justifying and sustaining the investment in research, development and infrastructure for large-scale, error-corrected quantum computing hinges on the community's ability to provide clear evidence of its future value through concrete applications. Further, in order to maintain technological momentum, it is critical to match investment growth and hardware progress with algorithmic capabilities. The time to discover quantum algorithms is now, and this provides an exciting opportunity where theory research can have enormous leverage and impact.

This perspective outlines the wide-ranging research required to create practical quantum computer applications. We divide that quest into five stages:
\begin{enumerate}
    \item Discovery of new quantum algorithms in an abstract setting.
    \item Identification of problem instances exhibiting quantum advantage.
    \item Establishing quantum advantage for a real-world application.
    \item Optimization, compilation and resource estimation for a use case.
    \item Application deployment.
\end{enumerate}
Along the way, we identify several key features of quantum algorithms research that we believe are crucial for delivering a useful real-world application. Our perspective focuses on the development of quantum applications realizing computational speedups, particularly those designed for fault-tolerant quantum computers. We set aside other important areas of quantum advantage---e.g., in sensing, metrology, or communication---to concentrate on practical computational applications. (For a recent review on categories of quantum advantage, see \cite{Huang2025Vast}).

In the following subsections, we briefly introduce several prominent theses of our perspective, which will be discussed in more depth throughout the article. We then discuss the five stages categorizing the maturity of a quantum application, key challenges for each stage, and how we can make progress. We include standalone sections on the importance of verifiability, on quantum simulation, compilation in the early-fault tolerant regime, and the economics of quantum application discovery.

\subsection*{Focus on verifiability}

It is important to distinguish between quantum algorithms that could one day provide the basis for a practically relevant computation, and those that will not. We argue that in the real world, computations are not useful unless they are verifiable in the sense that the quality of the solution can be efficiently checked or measured. If we can efficiently spoof the output of a quantum computation with no efficiently detectable change in performance, then such spoofing would be easier than building and running a quantum computer. If spoofing the output affects the performance in an efficiently detectable way, then we should be able to use that to verify the computation.

We believe that verifiability is a necessary (but insufficient) condition for the utility of quantum algorithms primitives. This rules out quantum advantages based on sampling from a scrambled quantum state \cite{Aaronson2011Computational, Arute2019Quantum}. While such tasks represent interesting demonstrations of the power of quantum computing and lower-bound the resources required for quantum advantage in more useful contexts, if one cannot efficiently measure how well they are solving the problem at hand, the real-world impact of that solution would also be inefficient to measure.

Efficient classical verification is the highest standard. However, we can set a lower bar for assessing the potential utility of quantum algorithms in quantum simulation: the output of a useful quantum algorithm should at least be verifiable by another quantum computer. This allows for cross-verification between quantum devices or for verifying the computation against nature itself. For instance, consider a quantum simulation algorithm that computes a physical observable. If two different quantum computers could run the simulation and get the same answer, one can at least regard this as an experimentally testable prediction about the world. This can not be said for some sampling tasks such as random circuit sampling. We note that Refs.~\cite{QuantumFrontiers2025,Lanes2025Framework} have also recently called attention to the importance of verification.

\subsection*{Focus on finding hard instances and ensembles}

Many research papers in quantum algorithms fail the following litmus test: if given a quantum computer tomorrow, could the quantum algorithm be implemented and realize a quantum advantage? In order to do this, one needs not only a quantum algorithm but also a way to sample a quantumly-easy yet classically-hard instance. Many quantum algorithms come with exponential speedups in query models or worst case speedup guarantees such as BQP-Completeness. However, most papers lack a method to actually generate instances with quantum advantage in a computational model aside from embedding other, already known-to-be-hard problems with an advantage (e.g., factoring).

In quantum simulation there appear to be a wealth of systems that folklore holds to be intractable for classical computers yet tractable for quantum computers. However, the community does not have many candidates for verifiable quantum simulation problems with an asymptotic quantum speedup that is convincingly established in the average case. We hope that this presents an exciting new direction for quantum algorithms researchers, one which is impactful and promising yet relatively underexplored. We note that other recent perspectives have also called attention to this specific challenge \cite{QuantumFrontiers2025, Huang2025Vast, buhrman2025formalframeworkquantumadvantage}.

The most important reason to seek knowledge of instances with a quantum speedup is because an understanding of the type of structure that leads to advantage can help to find real-world problems where the advantage persists. Such problems are also “shovel-ready” for hardware demonstrations of quantum advantage. Finally, we discuss that if the problem has the property that a classical machine learning method would struggle to generalize from training on a small number of quantumly computed solutions, this would prevent the eventual obsolescence of quantum computing in the context of that problem.

It should be noted that a core tenet of our approach, shared by many others in the field, is that achieving practical quantum advantage in computational models in the foreseeable future (i.e., anytime in the next two decades) likely requires super-quadratic speedups \cite{Babbush2021Focus, Hoefler2023Disentangling}. While many problems are amenable to quadratic speedups using standard techniques like amplitude amplification, discovering algorithms with higher-order speedups is considerably more difficult. Henceforth, when we refer to a computation as having quantum ``advantage'' we mean a computation run at such a scale that it actually requires fewer computational resources (e.g., time) than classical competition. By contrast, when we write quantum ``speedup'', we are referring to the algorithm scaling.

\subsection*{Focus on connecting to real-world applications}

Even a quantum algorithm that is efficient for explicit classically-hard instances is not yet a quantum application. Perhaps the most significant bottleneck in delivering value is the translation of abstract algorithms into solutions for real-world problems where a practical advantage holds under all physical and economic constraints. Outside of quantum simulation and cryptanalysis, this connection has proven difficult to demonstrate. This ``application search'' is often undervalued in the academic community compared to the discovery of new abstract algorithms, creating an imbalance: the field has a growing portfolio of theoretical tools but a scarcity of demonstrated, practical use cases.

This challenge stems from two primary difficulties. The first is technical: most algorithms with super-quadratic speedups come with a long list of stringent criteria—such as requirements on data structure, matrix sparsity, or condition number—that are rarely met in problems of commercial interest. The second is sociological: a persistent knowledge gap exists between quantum algorithmists and real-world domain specialists. Finding a new application requires a rare, cross-disciplinary skill set to bridge this divide. We argue that a ``problem-first'' approach, where one starts with an industrial challenge and tries to invent a quantum solution, though attempting to more directly address an end-user's needs, is rarely fruitful. A more effective strategy is the ``algorithm-first'' approach: begin with a known quantum primitive (e.g., quantum simulation) that offers a clear advantage, and then search for real-world problems that map onto the required mathematical structure. We posit that generative artificial intelligence might be a useful tool for helping to bridge the knowledge gap between disparate fields.

\subsection*{Focus on architecture in early fault-tolerance}

An important final step in developing quantum applications is the compilation of logical circuits and the optimization of the resources required for realization within error-correcting codes. Much has been written lately about the dawn of the “early fault-tolerant” era and an emerging important direction in the field will be to design algorithms for this era. However, in our section on early fault-tolerant compilation, we caution that simple rules of thumb such as “focus on reducing circuit depth and the number of logical qubits” are often insufficient (and sometimes actively harmful) for guiding the search for early fault-tolerant algorithms.

Just as success in the era of noisy intermediate scale quantum devices required careful consideration of the details of the target hardware platform, we argue that success in the early fault-tolerant era demands careful consideration and perhaps even co-design of the quantum error-correction architecture. While it is not practical for all algorithmists to program at the level of stabilizer diagrams, we argue that algorithmists familiar with intermediate representations and cost models will provide unique capabilities. However, these models depend sensitively on certain assumptions of the underlying architecture that still need to be developed and optimized.

\begin{figure*}[t]
    \centering
    \includegraphics[width=0.8\textwidth]{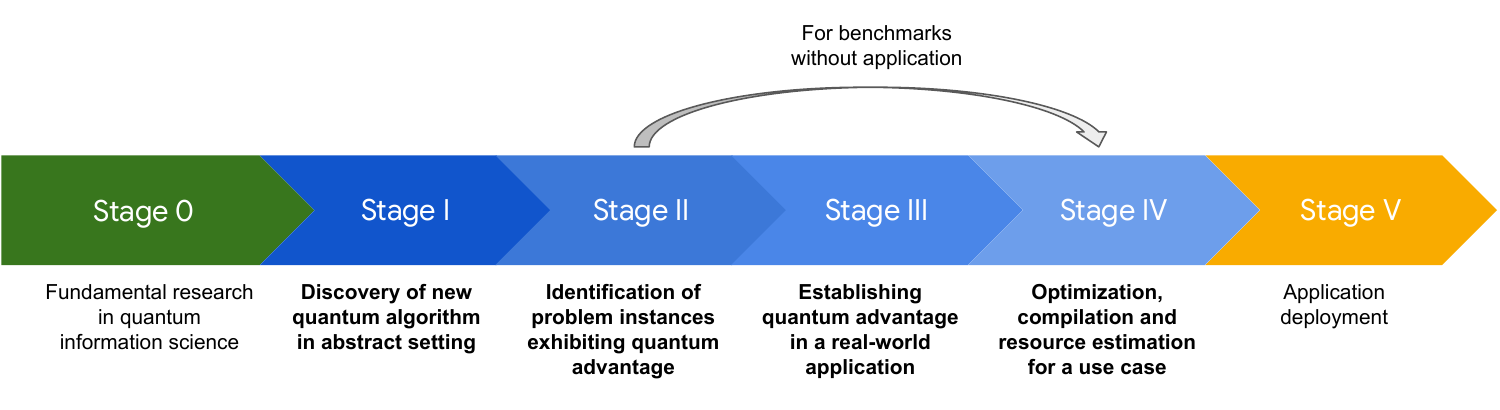}
    \caption{The stages of quantum application development. This framework is a conceptual model of application research maturity and progression is not always linear. Some seminal works, like Shor's factoring algorithm, addressed multiple stages at once (I, II, and III), while other research paths may skip stages entirely. For example, research into ``tests of quantumness'' often focuses on minimizing resource requirements tasks with quantum advantage irrespective of practical applications. Ultimately, this framework serves to categorize research and identify the key bottlenecks on the path to realizing practical quantum applications.}
    \label{fig:stages}
\end{figure*}

\section*{The five stages of quantum applications research}

We categorize the development of quantum applications into five stages, as depicted in \Cref{fig:stages}. In this section, we provide a high-level summary of these stages. The remainder of the article will then offer a more detailed discussion of the challenges inherent in each.

\paragraph*{Stage 0 --- Fundamental research in quantum information science.}

Stage 0 encompasses the foundational work in quantum information and complexity theory that guides the search for quantum speedups. This research often establishes fundamental limits on the power of quantum computation. For example, classical simulation algorithms \cite{Gottesman1998Heisenberg}, quantum query lower bounds \cite{Beals2001Quantum}, and no-go theorems (e.g. the ``no fast-forward theorem'' \cite{Berry2007Efficient, Atia2017Fast}) can prove that no quantum speedup is possible for certain problems, steering researchers away from dead ends. Likewise, there are many examples of quantum complexity theory teaching us that certain problems (e.g., preparing sufficiently low temperature Gibbs states of many Hamiltonians \cite{Chen2023Quantum, Chen2024Local}) are sufficiently difficult in the worst case (e.g., QMA-Hard \cite{Kitaev2002Classical}) that we should not expect efficient quantum algorithms that work for all instances. Such results help quantum algorithmists focus their attention and effort on more realistic goals. 

Stage 0 research is typically abstract, curiosity-driven, and conducted primarily within academia rather than industrial groups. While critically important, since Stage 0 is not algorithm or application development per se, we will not discuss it further in this perspective.

\paragraph*{Stage I --- Discovery of new quantum algorithm in abstract setting.}

Stage I involves the discovery, design and analysis of new quantum algorithms. In Stage I, the goal is often to introduce a new quantum computational primitive and to give some sort of evidence for how it performs. Work at this stage often focuses on worst case asymptotic complexity, concerned with the scaling of the algorithm for the most difficult possible instance. Algorithms that invoke blackbox oracles to solve problems in simplified computational models such as the query model would be in scope for Stage I. This stage includes many foundational results of the quantum algorithms field. Canonical examples include Grover's algorithm \cite{Grover1996Fast}, Simon's algorithm \cite{Simon1997Power}, quantum phase estimation \cite{Kitaev1995Quantum} and quantum algorithms for the Abelian hidden subgroup \cite{Kitaev1995Quantum}. It also encompasses broad frameworks like the quantum adiabatic algorithm \cite{Farhi2000Quantum}, the quantum approximate optimization algorithm \cite{Farhi2014Quantum}, and quantum signal processing \cite{Low2016HamSim,Gilyen2018singular}, as well as algorithms for BQP-complete problems such as the solving systems of linear equations \cite{Harrow2009Quantum} or estimating the Jones polynomial \cite{Aharonov2006Polynomial}.

\paragraph*{Stage II --- Identification of problem instances exhibiting quantum advantage.}

A Stage I result advances to Stage II when one is able to efficiently generate concrete problem instances (even if contrived) where a quantum computer is expected to outperform the best-known classical algorithms. For example Stage II can be accomplished when a problem is shown to have an average-case super-quadratic speedup across an efficiently constructible ensemble. This is a crucial step towards practical utility. Results only demonstrating worst-case speedup are not yet Stage II since one might not know how to generate classically difficult instances except by embedding other problems that were already known to have a quantum speedup (e.g., factoring).

Sampling problems with large quantum speedups seem to be abundant \cite{Boixo2018Characterizing, Aaronson2011Computational}. However, we will argue that Stage II algorithms with potential for utility should be either classically or quantumly verifiable and it has been considerably more difficult to advance problems with that property to this stage. Examples of quantum algorithms papers that have addressed Stage II for decision problems include quantum algorithms for Gauss sums \cite{VanDam2002Efficient}, tensor principal component analysis \cite{Hastings2020Classical}, the optimal polynomial intersection problem \cite{Aharonov2006Polynomial} and the simulation of higher order Out-of-Time-Order Correlators (OTOCs) on random circuits \cite{Google2025}. (See Ref.~\cite{king2025simplifiedversionquantumotoc2} for a formal definition of the OTOC problem.)

Stage II results must confront two related challenges. The first is to identify a problem instance such that the output of the quantum algorithm avoids concentration on an average input-independent value, and therefore reveals an input-dependent signal. The second challenge is the need to consider a vast array of (possibly heuristic) classical algorithms that could compete with the quantum algorithm. Beyond being a necessary step toward applications, our belief is that Stage II is as much the heart of the quest for quantum advantage as Stage I.

\paragraph*{Stage III --- Establishing quantum advantage in a real-world application.}

Stage III is the critical step of connecting an abstract quantum advantage from Stage II to a specific, real-world application. A result reaches this stage when it identifies a practical problem where a quantum algorithm is expected to deliver a meaningful speedup, after accounting for all real-world constraints and overheads. This stage is challenging because the ``fine print'' of many quantum algorithms often prevents a direct mapping to practical problems \cite{Aaronson2015Read}.

The most famous Stage III success is Shor's algorithm \cite{Shor1999Polynomial}, which had the immediate application of breaking widely used cryptosystems \cite{Litinski2023How, Garn2025Quantum}. Since then, quantum simulation has seen the most progress. Our team has developed Stage III case studies detailing how quantum simulations could be applied to challenges in fusion energy, battery design, and drug discovery \cite{Rubin2024Quantum, Rubin2023Fault, Santagati2024Drug, Berry2024Quantum, vonBurg2021Quantum, Caesura2025Faster}. Outside of quantum simulation and cryptography, directly linking novel algorithms to high-value practical problems in a manner that retains a quantum speedup remains a major outstanding goal for the field.

\paragraph*{Stage IV --- Optimization, compilation and resource estimation for a use case.}

Stage IV shifts the focus from theoretical speedups to the practical engineering challenges of optimization, compilation, and resource estimation for a specific use case. This marks a significant departure from earlier stages: the emphasis moves from asymptotic analysis to concrete, finite resource costs, such as the exact number of qubits and gates required. This work meticulously analyzes and optimizes the constant factors determining an algorithm's real-world viability. Canonical examples include the long series of papers that have systematically reduced the estimated resources for simulating quantum chemistry \cite{Low2025Fast} or breaking RSA encryption \cite{Gidney2025Factoring}. More broadly, Stage IV encompasses the development of compilers for mapping logical circuits onto error-correcting codes, as well as the creation of software tools, like Qualtran \cite{Harrigan2024Expressing}, Q\# \cite{Singhal2022Q}, QREF \cite{PsiQ_QREF}, Bartiq \cite{PsiQ_Bartiq}, Qrisp \cite{Seidel2024Qrisp}, and Silq \cite{Bichsel2020Silq}, designed (at least in part) to automate and streamline these painstaking resource estimates.

\paragraph*{Stage V --- Application deployment.}

Stage V represents the final phase: deploying a proven quantum solution into a practical, real-world workflow. As of today, this stage is entirely prospective, as no quantum application has yet been implemented in hardware with a conclusive advantage on a problem of real-world consequence. 
Importantly, the motivation for deployment will be agnostic to the fact that the underlying technology is quantum, relying entirely on the fact that it is a better solution to the problem or application at hand.
Perhaps the closest current example is certified random number generation \cite{Aaronson2023Certified}, though its impact to-date remains limited.
Stage V will likely involve creating industry-specific application programming interfaces, environments and workflows---for example, a cloud service that allows clients to seamlessly offload tasks to a quantum computer. Although this stage is certain to have significant challenges associated with it, we elect not discuss it further here since most of this work is in the future.

\section*{Stage I research}

While discovering new algorithms is challenging, Stage I has seen significant progress in recent years, with active research in areas including quantum machine learning, Gibbs state preparation, and quantum algorithms for differential equations among other areas \cite{Huang2022Quantum, Huang2023LearningMany, Huang2023Learning, Huang2024Learning, Allen2025Quantum, Yamakawa2024Verifiable, Chen2023Quantum, Jordan2025Optimization, Babbush2023Exponential, somma2025quantum, Hastings2018Short, Lloyd2016Quantum}. See \Cref{tab:stage1} for a non-comprehensive but representative list of some Stage I results with super-quadratic speedups.

\begin{table*}[t]
\centering
\begin{tabular}{ll}
\toprule
\textbf{Type} & \textbf{Result} \\
\midrule
\multirow{5}{*}{BQP-complete} & Computing knot invariants (Jones polynomials) \cite{Aharonov2006Polynomial, Aharonov2011BQP, Shor2007Estimating} \\
 & Solving linear systems of equations \cite{Harrow2009Quantum} \\
 & Coupled classical oscillator dynamics \cite{Babbush2023Exponential} \\
  & Preparing local minima of quantum Hamiltonians \cite{Chen2024Local} \\
 & Universal quantum simulation \cite{Lloyd1996Universal} \\
\midrule
\multirow{5}{*}{Oracular} & Simon's problem \cite{Simon1997Power} \\
 & Forrelation \cite{Aaronson2015Forrelation} \\
 & Yamakawa-Zhandry \cite{Yamakawa2024Verifiable} \\
 & Glued trees \cite{Childs2003Exponential} \\
 & Abelian hidden subgroup \cite{Kitaev1995Quantum} \\
\midrule
\multirow{5}{*}{Other} & Adiabatic algorithm \cite{Farhi2000Quantum, Aharonov2008Adiabatic, Gilyen2021Sub} \\
 & Early results on QAOA \cite{Farhi2014Quantum} \\
 & Topological data analysis \cite{Lloyd2016Quantum} \\
 & Simulating differential equations \cite{Costa2019Quantum, Berry2014High, Liu2021Efficient,Low2024Eigenvalue} \\
  & Generative learning of shallow quantum circuits \cite{Huang2024Learning, Huang2025Generative} \\
\bottomrule
\end{tabular}
\caption{A representative (but not at all comprehensive) list of some Stage I results with super-quadratic speedups.}
\label{tab:stage1}
\end{table*}

However, this does not mean Stage I is without challenge. We observe that despite a dramatic increase in the size of the quantum computing field throughout the last decade, a smaller percentage of researchers today seem to be working on certain foundational topics in algorithms than in the past. For example, the study of quantum algorithms with algebraic structure, such as for the Hidden Subgroup Problem, has become a sparsely populated subfield. This is despite a history of profound results and a sense among practitioners that more discoveries await \cite{Kuperberg2005Subexponential, Friedl2014Hidden, Hallgren2007Polynomial, VanDam2006Quantum, Childs2014Constructing, VanDam2008Classical, VanDam2002Efficient, Kedlaya2006Quantum}. This trend is particularly concerning in the context of post-quantum cryptography. The security of such schemes can only be established by a robust community effort to find quantum attacks. Yet, only a small fraction of the few researchers with the rare dual expertise in quantum algorithms and cryptography are actively working on this. This is alarming because a critical vulnerability discovered after post-quantum cryptography has been widely deployed would be catastrophic.

That a relatively small fraction of quantum algorithms researchers appear focused on foundational, high-risk research in algorithms might be explained in part by structural incentives in academia and industry that inadvertently encourage risk-averse behavior. Finding ways of encouraging more researchers to pursue bold, foundational searches for new quantum algorithms would meaningfully expand the frontier of what our field can achieve. There is a large potential upside: a short paper with a genuinely new idea can redirect years of effort. Evaluation and funding mechanisms ought to reward originality, problem-finding, and careful exploration of hard ideas (including well-documented negative results), not only incremental improvements, which would empower more researchers to aim for breakthroughs.

Our experience suggests that new quantum algorithms rarely appear from a vacuum and instead still find some inspiration from prior (often ``Stage 0'') work. Even Shor’s algorithm was inspired by the period finding of Simon’s algorithm. In turn, Shor’s algorithm inspired subsequent work in period finding, phase estimation, and algorithms for the Abelian hidden subgroup problem. Many recent Stage I innovations from our own team at Google have followed a pattern of generalization and modification. For instance, our work on planted inference \cite{Schmidhuber2025Quartic} was a generalization of prior results on tensor principal component analysis (PCA) \cite{Hastings2020Classical}. Our algorithms for simulating classical oscillators \cite{Babbush2023Exponential}, and the subsequent work on Shadow Hamiltonian simulation \cite{Somma2025Shadow} and linear matrix equations \cite{somma2025quantum}, were inspired by and extended previous quantum algorithms for differential equations \cite{Costa2019Quantum, Berry2014High, Liu2021Efficient}. Our recent work on decoded quantum interferometry (DQI) \cite{Jordan2025Optimization} bore substantial resemblance to prior work on quantum reductions \cite{Aharonov2003Adiabatic, Regev2004Quantum, Aharonov2005Lattice, Regev2009Lattices} and oracle separations \cite{Yamakawa2024Verifiable}. This process---studying a speedup or interesting quantum subroutine in a specific context and then adapting its core mechanism to solve a different class of problems---has been a reliable engine for Stage I discovery.

\subsection*{Verifiability or reproducibility as necessary conditions for utility}

How can we tell if an abstract quantum algorithm has the potential for utility in applications? This section proposes a necessary (but insufficient) condition as such a filter: useful quantum computations must be verifiable, or at least, ``quantumly verifiable''.

We consider an output to be verifiable if a quantum or classical computer can efficiently check if the answer is correct (without requiring quantum communication). We argue that we cannot consider a solution to be useful if there is no way to verify it in a reasonable time. If the quality of the solution makes any difference in the real world, then the real world is effectively giving feedback on the quality of the answer and this feedback could be used for verification. For example, imagine using a quantum computer to optimize a technology. If the new design it suggests performs measurably better, that real world success would be an informal verification of the result. It follows that tasks of relevance to the classical world ought to be classically verifiable in order to be useful.

An output is considered quantumly verifiable if the output can be verified by a trusted quantum computer. For example, any computation that is independently \emph{reproducible} will be quantumly verifiable: running the same quantum algorithm across different quantum devices yields consistent results. Here we are referring to the final answer output by the quantum algorithm (e.g., the estimate of the expectation value or the final factors rather than the exact samples or repeated runs that were required to determine that answer). This is crucial for quantum simulations because their purpose is to make predictions about nature (which can be imagined as performing a quantum computation). For example, if a quantum computer helps design a new superconductor, the simulation would be confirmed if the material is created and observed to have the predicted properties.

A task like random circuit sampling fails the conditions for utility. It is believed that the output of random circuit sampling is not efficiently verifiable or reproducible, even by another quantum computer. Of course, many sampling tasks are computationally useful. For example, Monte Carlo sampling algorithms are widely used in practice. However, applications of sampling typically use samples to extract meaningful information or specific features of the underlying distribution (e.g., Monte Carlo is often used to evaluate integrals). Sampling from generative AI models is only useful if the samples can be evaluated under some quality metric. In contrast, samples obtained from random quantum circuits lack any efficiently discernible features. If a quantum algorithm generates samples containing meaningful signals from which one could extract classically hard-to-compute values, then the problem ultimately being solved is not sampling but a quantumly verifiable task such as expectation value estimation with sampling as a subroutine.

Quantum computers first obtained quantum advantage in sampling against the best classical algorithms in 2019 \cite{Arute2019Quantum}. By contrast, the first claims of quantum advantage in computing an expectation value to have withstood preliminary efforts at classical simulation seem to have occurred only in the last year. Notable claims were made by IBM \cite{Kim2023Evidence} and D-Wave \cite{King2025Beyond} in 2024 but then challenged by several groups shortly thereafter using relatively standard methods \cite{Begusic2024Fast, Kechedzhi2024Effective, Tindall2024Efficient, Tindall2025Dynamics, Mauron2025Challenging}. This year, claims were made by Quantinuum \cite{Haghshenas2025Digital} and Google \cite{Google2025} and the Quantinuum results have recently been simulated classically \cite{Mandra2025}. Thus, even though these recent Stage II demonstrations are not yet solving practical problems, we believe that solving expectation value problems with quantum advantage as opposed to sampling problems represents clear progress towards more useful quantum computing.

Finally, we emphasize that we consider these conditions as guiding principles for assessing the utility of quantum algorithms, rather than as an absolute criterion. Certain computations, such as certified random number generation \cite{Aaronson2023Certified}, may remain useful even when verification is inefficient, although such cases generally lie outside the computational model considered in this work.

\begin{table*}[t]
\centering
\begin{tabular}{llll}
\toprule
\textbf{Stage I inspiration \hspace{1cm}} & \textbf{Problem} & \textbf{Speedup} & \textbf{Verifiable} \\
\midrule
\multirow{5}{*}{Period finding \cite{Simon1997Power}} & Factoring and discrete log \cite{Shor1999Polynomial} & Exponential & Classically \\
 & Principal ideal and Pell’s equation \cite{Hallgren2007Polynomial} & Exponential & Classically \\
 & Zeta function of algebraic curves \cite{Kedlaya2006Quantum} & Exponential & Quantumly \\
 & Unit and class groups \cite{Hallgren2005Fast} & Super-polynomial\qquad \quad  & Classically \\
  & Two-variable exponential congruences \cite{VanDam2008Classical} & Cubic & Classically \\
\midrule
\multirow{4}{*}{Phase estimation \cite{Kitaev1995Quantum}} & Gauss sums \cite{VanDam2002Efficient} & Super-polynomial & Quantumly \\
 & Tensor principal component analysis \cite{Hastings2020Classical} & Quartic & Classically \\
 & Sparse learning parity with noise \cite{Schmidhuber2025Quartic} & Quartic & Classically \\
&  Community detection \cite{Schmidhuber2025Community} & Quartic & Classically \\
\midrule
\multirow{2}{*}{Abelian hidden shift \cite{Kuperberg2005Subexponential}} & Legendre symbol \cite{VanDam2006Quantum} & Exponential & Classically \\
 & Elliptic curve isogenies \cite{Childs2014Constructing} & Super-polynomial & Classically \\
\midrule
\multirow{2}{*}{Regev’s reduction \cite{Regev2004Quantum}} & Optimal polynomial intersection \cite{Jordan2025Optimization,Gu2025Algebraic,Hillel2025Optimization} & Exponential & Classically \\
 & Yamakawa-Zhandry (instantiated) \cite{Yamakawa2024Verifiable, Briaud2025Quantum} & Exponential & Classically \\
\midrule
Quant.~approx.~opt.~algo. (QAOA) \cite{Farhi2014Quantum} & QAOA for 8-SAT \cite{Boulebnane2024Solving} & Super-quadratic & Classically \\
\midrule
Random circuit sampling (RCS) \cite{Boixo2018Characterizing} & RCS out-of-time-order correlators \cite{Google2025} & Exponential & Quantumly \\
\bottomrule
\end{tabular}
\caption{The short list of Stage II results that are currently known to the authors of this perspective for which there is an exponential-size ensemble with an average-case super-quadratic quantum speedup. This is in contrast to problems such as FeMoco or the Hubbard model where there is effectively a finite number of distinct calculations we expect to have quantum advantage. Also worth mentioning but not included in this table is work on proofs of quantumness based on trapdoor claw-free functions where verification is only efficient using hidden side information~\cite{brakerski2021cryptographic,kahanamoku2022classically,brakerski2020simpler,arabadjieva2024single,zhu2023interactive}.}
\label{tab:stage2}
\end{table*}

\section*{Stage II research}

We consider a quantum algorithm to have reached Stage II only when one can generate concrete instances of a problem that are expected to exhibit verifiable quantum advantage. For example, this would be satisfied if one identifies an ensemble of quantumly verifiable problem instances that can be efficiently sampled for which a quantum algorithm gives a super-quadratic speedup over the best known classical algorithms, with probability vanishing at most polynomially in the problem size (e.g., they are average-case hard classically). While this is a demanding criterion, several results have met it. Examples of papers that have addressed Stage II include Shor's algorithm \cite{Shor1999Polynomial}, the DQI algorithm applied to optimal polynomial intersection (OPI) \cite{Jordan2025Optimization}, algorithms for tensor PCA \cite{Hastings2020Classical} and speedups for other number-theoretic problems like Pell's equation \cite{Hallgren2007Polynomial}. One could also consider using a quantum computer to efficiently generate or sample hard instances, although we are not aware of any examples of this kind.

It is acceptable for Stage II results to target abstract or even contrived problems. Trying to contrive examples with advantage can be a good first step towards later applications because if a quantum advantage cannot be demonstrated in an idealized, hand-picked setting, the prospect of discovering one in a complex real-world application (Stage III) is probably negligible. Despite its critical role as a bridge from theory to practice, Stage II is frequently overlooked by the quantum community, and its difficulty is often underestimated. However, those working at the intersection of quantum computing and cryptography often do focus on Stage II results because cryptosystems based on computational security usually require average case notions of problem complexity.

Many important algorithms remain in Stage I because their advantage has only been proven in abstract settings that do not allow one to generate concrete problem instances with quantum advantage. Consider results in the query model, such as Simon's algorithm \cite{Simon1997Power} or the welded trees problem \cite{Childs2003Exponential}. While they demonstrate a proven separation from classical computing, the advantage relies on a black-box oracle. Without a method to instantiate these oracles in a way that remains classically hard, they have not addressed Stage II. Similarly, a paper describing an algorithm that provides only a worst-case speedup has also failed to address Stage II. The initial work on QAOA \cite{Farhi2014Quantum}, for instance, was a landmark Stage I achievement that showed a better worst-case approximation ratio than was known classically at the time. However, this worst-case guarantee did not allow one to sample problem instances with quantum advantage (even before better classical algorithms were introduced). In contrast, subsequent numerical studies identifying a scaling advantage for QAOA on particular problem classes \cite{Boulebnane2024Solving} made a successful push into Stage II.

A common misconception is that proving BQP-Completeness implies a practical promise of quantum speedup. However, we argue that on its own, BQP-Completeness is not just a hallmark of Stage I but a suggestion that more work must be done to get to Stage II. It indicates one has found a new universal language for quantum computation, rather than new problem instances with demonstrable advantage. Of course, one can always derive ensembles of BQP-Complete problems with average-case quantum speedups by reducing already-known Stage II+ results (e.g.~factoring) into the new framework but that does not help to elucidate new, natively hard instances. The issue is that proving a problem BQP-Complete does not inherently specify a distribution of hard inputs. Furthermore, ``natural'' ensembles of instances of BQP-Complete problems---like sparse random matrices for the HHL algorithm \cite{Harrow2009Quantum}, random braids for knot invariants \cite{Aharonov2006Polynomial} or the dynamics of random oscillator networks \cite{Babbush2023Exponential}---are often classically easy on average. This occurs when quantities of interest concentrate to classically predictable values, eliminating any potential for quantum advantage. Constraining problems in a way that precludes BQP-Completeness may in some cases help to avoid this concentration phenomena.

A key distinction among Stage II results is their long-term vulnerability to being solved by an efficient classical machine learning algorithm given a finite amount of classical data generated by the quantum computer. This allows us to separate Stage II results into two groups.

The first category contains problems where there are a finite number of distinct instances with a quantum advantage. This includes important simulation targets like the FeMoco molecule or the phase diagram of the Hubbard model, which have a limited number of distinct phases or parameters. Here, a quantum computer might only be needed to provide the initial data. It's conceivable that after solving enough instances, a classical model could learn the underlying patterns and generalize to the remaining problems, reducing the need for further quantum computation. For example, in electronic structure force-field calculations one can already use a few high accuracy data points calculated using expensive (classical) methods to boost the accuracy of data taken with cheaper, less accurate methods \cite{Unke2021Machine}. This first class of problems is related to what complexity theorists call the complexity class P/Poly: those that can be solved in polynomial time given a polynomial amount of ``advice''. (When this class is restricted using efficiently computable data from a polynomial number of quantum computations it is known as BPP/Samp.) Of course, one still requires a quantum computer to provide initial data, and understanding generalization, or lack thereof, remains a matter of ongoing research \cite{Huang2021Power, Huang2022Provably}.

The second category consists of problems where there is an exponentially-large ensemble of instances that are classically hard in the average case and the solution of any polynomial sized set of problem instances provides meaningful information about only a vanishingly small fraction of other problem instances.  Such fundamentally unlearnable structure implies a lack of regularity in the solution space of the problem in mapping from inputs to outputs, and the implication for the available application space is worthy of further study. For these problems, the advantage is likely to be permanent. Even after a quantum computer solves many examples, a classical machine-learning model would still be unable to generalize from that data to efficiently solve new instances. We list all results of this type known to us in \Cref{tab:stage2}. The second category of problems offers a potential unending need for quantum computers and value here might be more defensible for the businesses offering a quantum computer. However, the first category still forms a valuable and important application of quantum computing.

A core difficulty of Stage II arises from a fundamental tension between two competing forces. On one hand, a quantum algorithm must generate highly complex states with large entanglement to be difficult for classical computers to simulate. On the other hand, these same complex states often exhibit scrambling or thermalization, causing their observables to become statistically predictable. For a scrambled quantum state, observable values tend to concentrate tightly around their average values, erasing the unique ``signal'' associated with the specific problem input. A trivial classical algorithm can then compete by simply outputting this predictable average value. This tension appears to be a typical behaviour of randomly-sampled quantum states \cite{Bremner2009Are, Gross2009Most}, and states produced by the output of random quantum circuits \cite{Angrisani2024Classically}. Similarly, Hamiltonians of generic strongly correlated systems often thermalize to universal behaviors, with ensemble features predictable by random matrix theory \cite{Mehta2004Random}, or from single non-random quantum states \cite{Cotler2023Emergent}. A major challenge in Stage II is to find problem classes where corrections to this universal behavior are observable (i.e. not exponentially small), and not predictable by classical algorithms.

There are various rules of thumb for what make problems in many-body physics hard; e.g.: disorder, systems near phase transitions, properties at low but non-zero temperature, non-equilibrium phenomena, and glassy or topologically non-trivial phases. However, significant work is required to confirm that such properties lead to instances that do not fall prey to the aforementioned issues, yielding problems that could fall into \Cref{tab:stage2} above. Such results would give a better idea of asymptotically promising targets for quantum computers within many-body physics. Moreover, this search will likely expand and better our knowledge of interesting physical phenomena: previous work from the quantum computing community on many-body scars \cite{Bluvstein2021Controlling}, random circuits \cite{Boixo2018Characterizing}, and higher-order out of time-ordered correlators \cite{Google2025} have all opened new areas of research within the strongly-correlated physics community. Understanding the boundary where many-body systems become computationally complex can be viewed as a fundamental pursuit in the study of highly entangled quantum systems.

Another major challenge of Stage II is that the classical competition is the constantly improving ``moving target'' of practical, state-of-the-art heuristics, rather than the provable, worst-case algorithms that are often the point of comparison in Stage I. This creates a fundamental asymmetry. Today, we typically rely on theoretical proofs or limited simulations to assess a quantum algorithm's average-case performance, while classical approaches are benchmarked by their best-observed performance in practice. History shows that the power of many of the most impactful classical algorithms---including the Metropolis algorithm, the simplex method, and modern neural networks---were difficult to assess analytically and required large-scale empirical testing for their development \cite{Spielman2004Smoothed, Roughgarden2019Beyond}. This precedent suggests that the true potential of certain quantum heuristics may likewise remain hidden until we can explore them on scaled-up quantum hardware. Moreover, central Stage II questions such as the extent to which expectation values concentrate in output states would be straightforward to study using large quantum computers. Hence, while the classical moving target makes Stage II difficult today, future hardware may make it easier to identify promising avenues for quantum advantage.

Even without an immediate real-world application, Stage II results are valuable because they provide a concrete recipe for generating hard test cases, essential for validating hardware progress and demonstrating algorithmic milestones. This is the mentality underlying research into algorithmic ``tests of quantumness'', which treats Stage II as an end in and of itself. The goal is to set aside immediate practicality and instead answer the question: what is the simplest possible problem that demonstrates quantum advantage with the absolute minimum resources? We believe that the leading proposals to date (in the sense they solve classically intractable problems with the fewest quantum resources) are Jacobi factoring \cite{Kahanamoku-Meyer2025Jacobi} for a classically verifiable problem and OTOC of random quantum circuits \cite{Google2025} for a quantumly verifiable problem. By seeking the cheapest possible quantum advantage, such work provides a heuristic lower bound on the resources that might be needed to give advantage in a useful application. Such benchmarks also help to distill the essence of quantum advantage in a fashion which may help to sharpen intuition about where to look for necessary structures in practical contexts.

We believe Stage II in an exciting opportunity for researchers. One reason is that it falls into a disciplinary gap; e.g., perhaps theorists drawn to the abstract nature of Stage I, view the numerical and heuristic components of Stage II as ``too applied'' while application-focused researchers see the push for contriving hard instances as an esoteric aim. Another possible reason is that parts of the community might not appreciate the difficulty, or importance of finding instances with average-case speedups within a problem known to admit worst-case ones. For example, many physicists seem to either believe that finding Stage II results is trivial in quantum simulation and that is why there aren’t many papers about it, or believe that the problem is very hard and that is why there aren’t many papers about it. By highlighting how few Stage II results have been substantiated, we hope to encourage more researchers to fill this critical gap. Stage II is not merely a prerequisite for useful applications -- it represents a fundamental question at the core of understanding the complexity of quantum physics.

\section*{Stage III research}

Outside of cryptanalysis and quantum simulation, the search for practical quantum applications has proven remarkably difficult. This gap represents the core challenge of Stage III research: taking an algorithm with a theoretical speedup and finding a real-world problem where that advantage holds true under all practical constraints. While a broader portfolio of algorithms from the earlier stages would certainly help, the current scarcity of Stage III applications highlights that connecting theory to practice is a major bottleneck in its own right.

The two primary successes of Stage III, cryptanalysis and quantum simulation, are both exceptional cases where the path from algorithm to application was unusually direct. Quantum algorithms for factoring and discrete logarithms were immediately connected to breaking (Rivest–Shamir–Adleman) RSA and elliptic curve cryptography (ECC), respectively, and quantum simulation benefits from a natural mapping between problem and computer. (We discuss quantum simulation applications further in the subsequent section.) An important goal is to identify clear real-world applications of quantum computers outside of these two areas. Current levels of investment in quantum computing anticipate eventually unlocking a total addressable market well in excess of tens of billions of dollars per year and by the time large quantum computers are built, the commercial value of breaking legacy cryptosystems might be relatively limited. Investors are already quite accustomed to hearing that quantum computers will help with simulation and cryptanalysis and thus, compelling use case stories in other areas have significantly more marginal value in their ability to inspire and incentivize investment.

A significant concern is the relative lack of academic attention devoted to Stage III compared to the earlier stages. Within the research community, Stage I and II work is sometimes (in our view, unfairly) perceived as more ``fundamental'' and thus, more important than Stage III. A paper identifying a novel real-world use case for an existing quantum algorithm, for instance for linear systems \cite{Harrow2009Quantum} or topological data analysis \cite{Lloyd2016Quantum}, would likely not receive the same acclaim as the discovery of a new algorithm. This cultural bias creates a critical imbalance. The field has a growing portfolio of abstract algorithms but a severe scarcity of demonstrated, practical use cases---a gap that impedes the overall health of the field. Inspiring more researchers to tackle the challenges of Stage III was a primary motivation for our team's proposal and sponsorship of the Google Quantum AI XPRIZE Quantum Applications competition \cite{XPrizeQC}.

We posit that there are two basic reasons that Stage III is so challenging. The first (relatively obvious) reason is that Stage I and Stage II successes are rare and always come with a long list of criteria that must be met in order for super-quadratic quantum advantage to be possible. For example, quantum algorithms for solving linear systems $Ax=b$ \cite{Harrow2009Quantum} generally require $b$ to be prepared on the quantum computer with polylog($N$) resources, $A$ to be sparse and have polylog($N$) condition number, and these algorithms only enable sampling the entries of $x$ proportional to their magnitude. While the original paper on this topic (the ``HHL'' algorithm) has over 4k citations, we are not aware of any paper identifying a promising real-world instantiation with super-quadratic speedup, or even an ensemble of average-case hard instances placing it in Stage II. Additionally, many quantum algorithms pertain to problems that look somewhat esoteric, involving specific number theoretic or algebraic structure (e.g., \cite{Hallgren2007Polynomial, VanDam2006Quantum, Childs2014Constructing, VanDam2008Classical, VanDam2002Efficient, Kedlaya2006Quantum, Jordan2025Optimization}), and it can be difficult to find contexts in the real world where this structure naturally arises.

A second, purely sociological challenge for Stage III is the knowledge gap between quantum algorithmists and real-world domain specialists. Finding a new application is an ``application search'' that requires a rare, cross-disciplinary skill set to build bridges between abstract theory and practical problems. This work demands breadth over depth, which may be one reason it is often undervalued within traditional academic structures. In quantum simulation, many quantum researchers are physicists or even chemists with pre-existing domain expertise in classical many-body simulation, which effectively closed this knowledge gap from the start. In contrast, few quantum algorithmists have knowledge of fields like epidemiology or genomics, where algorithms for problems like topological data analysis \cite{Lloyd2016Quantum} might apply.

Some application discovery initiatives apply a ``problem-first'' strategy: ask a domain expert or customer for their hardest or most important problem, then try to design a quantum algorithm to solve it. This approach is rarely fruitful because quantum speedups tend to require exceptional mathematical structure. A more effective strategy is the ``algorithm-first'' approach. The search should begin with a known quantum primitive that offers an advantage and then search for real-world problems that map onto that structure. To facilitate this algorithm-first search for applications, our team and collaborators are creating a ``Compendium of Super-Quadratic Quantum Advantage''. This document will profile several dozen algorithms outside of simulation and cryptanalysis, each with a known super-quadratic speedup. The entries will give a concise description of the abstract problem and its ``regime of advantage'' in simple terms, without detailing the quantum algorithm. The compendium is designed to be readable by anyone with a technical undergraduate degree, assuming no prior knowledge of quantum computing. Our goal is to empower domain experts to identify potential applications in their own fields. If an expert finds a match between a problem in the compendium and a challenge they face, they can seek collaboration with quantum specialists to assess if the advantage translates to their use case.

We observe that this sociological knowledge gap makes Stage III work particularly well-suited for acceleration by generative AI. While AI is making impressive progress in advanced reasoning, our experience is that it is not yet broadly capable of the creative leaps required for most Stage I algorithm discovery. Part of its power today lies in leveraging a vast breadth of knowledge to connect disparate fields---the exact skill needed for the application search of Stage III. The agent's task would not be to invent, but to recognize: to scan its knowledge base for real-world problems that match the structure of known quantum speedups, even if they are described using different terminology in another field's literature. Our team has already seen early success with this approach using an internal Google tool based on Gemini \cite{GoogleResearchAI}.

The path through these stages of application maturity is not always linear. Sometimes, engaging directly with a Stage III application can provide the necessary guidance to solve the challenges of Stage II. A truly important real-world problem, like understanding a molecule with mysterious unresolved properties, often comes with a wealth of established classical research. Such work can provide a roadmap, highlighting the specific regimes where classical methods are known to fail and where phenomena like concentration are already well-understood. In such cases, the real-world application itself provides the context needed to satisfy certain Stage II criteria. Thus, while Stage III results should always at least build on established quantum primitives from Stage I, it may sometimes make sense to skip past Stage II to Stage III and matching to applications if doing so will help connect to literature about where the problem is actually difficult.

While it is important to proactively push Stage III forward, we conclude by noting that history teaches us to be optimistic. Graph theory, introduced by Euler in 1736, and the Fast Fourier Transform, discovered by Gauss in 1805, did not find much use until the advent of digital computing, where they became cornerstones. Number theory, an ancient art with no application whatsoever for millennia, became the bedrock of modern cryptography. The Paris Kanellakis prize given by the ACM is essentially a prize for ``Stage III'' results in pure mathematics, not unlike the Google Quantum AI XPRIZE Quantum Applications competition. The existence of this prize in a different field speaks to a universal challenge of bridging from theory to application.

\subsection*{Quantum simulation applications}

Aside from cryptanalysis, quantum simulation is the most mature domain for Stage III research, benefiting from a natural mapping between the problem and the computer. However, translating this potential into real-world impact still faces several significant hurdles. First, many powerful classical algorithms already effectively model a large number of systems that do not exhibit strong correlation with a significant correlation length \cite{Lee2023Evaluating, Chan2024Spiers}. Second, it can be difficult to establish that running a particular quantum computation (e.g. at a finite basis and computational cell size and with a certain model of the environment) will resolve the important open questions about the physics of a system of interest. Finally, a successful simulation must target the correct bottleneck in a larger technological process. In drug discovery, for instance, the primary bottleneck is not generating candidate molecules but assessing their efficacy and toxicity within the complex environment of the human body, a challenge currently solvable only by expensive, high-attrition clinical trials \cite{Scannell2012Diagnosing}. A quantum speedup not targeting the key bottleneck, therefore, might prove useful without being truly transformative.

A common proposal is to use quantum simulation for ``discovery''; it is easier to identify a classically intractable calculation than to know what one will learn from it. This is the case in both condensed matter physics and chemistry. In condensed matter, it is relatively simple to tune a model into a strongly-correlated, hard-to-simulate regime, but more difficult to know which specific calculation will reveal new physics. Similarly, in chemistry, the FeMo-cofactor of nitrogenase (FeMoco)---an enzyme involved in biological nitrogen fixation (fertilizer production)---is a famous target for quantum simulation. While there is hope that studying it will lead to better industrial processes, one cannot be certain of this without first learning what such simulations reveal.

Quantum chemistry has long been widely regarded as a leading candidate for real-world impact from quantum simulation. This is because chemistry is the central science at the heart of many industries and because \emph{ab initio} models of electronic structure aim to produce a high level of quantitative accuracy required for engineering applications of new molecules and materials. This contrasts with condensed matter physics, where models are usually designed to capture qualitative features---such as the phases of matter---rather than make precise numerical predictions about real systems. A computational chemist seeks to answer specific, quantitative questions: ``if I dope this battery cathode material with a particular atom, does its conductivity increase by 1\% or 3\%?'' A condensed matter physicist, by contrast, typically investigates broader, qualitative trends: ``what type of doping might improve conductivity?'' or ``is this a conductor or an insulator?''

While compelling molecular use cases like FeMoco \cite{Li2019Electronic} and P450 \cite{Goings2022Reliably} exist, they are relatively rare in the context of all industrially relevant chemistry. In contrast, promising solid-state applications are far more abundant, for several reasons \cite{Rubin2023Fault}. First, many key industrial targets such as heterogeneous catalysts involve solid-state materials, and their properties are often more directly determined by their electronic structure. That is, relative to molecular chemistry, fewer material properties depend strongly on entropic contributions that would need molecular dynamics to resolve. Second, it is easier to find classically intractable problems in solid-state systems. Many phases of matter in the solid-state have longer correlation lengths than molecules, creating complex electronic structure beyond the reach of classical simulation. This complexity, however, is a double-edged sword: the same properties that make these systems hard for classical computers also make them resource-intensive for quantum computers, demanding larger and more powerful devices \cite{Rubin2023Fault, Berry2024Quantum}.

Whether a true asymptotic, exponential quantum speedup exists for chemical ground states is a subject of legitimate debate. Some argue that ``natural'' molecules relevant to industry may have special properties that may make them solvable in polynomial time with advanced classical methods, like tensor networks \cite{Lee2023Evaluating, Chan2024Spiers}, even if many remain out-of-reach for practical methods today. Conversely, it's plausible that strongly correlated molecular ground states are not even efficiently preparable on a quantum computer in the true asymptotic limit. Yet another perspective is that the controllable errors in quantum algorithms provide a decisive advantage over classical methods where convergence guarantees may not even be feasible~\cite{vonBurg2021Quantum}.  

However, our view is that fixating on asymptotic scaling may be the wrong lens for assessing the practical value of quantum computing in chemistry. Molecules and the correlation lengths of materials are intrinsically finite; thus, an algorithm's utility depends on its performance on these specific instances, not its behavior at an infinite limit. Furthermore, while some classical methods may have favorable polynomial scaling in theory, they can be completely impractical for high-precision chemistry in practice. For example, the projected entangled pairs state (PEPS) method (the basis for claims that classical methods may efficiently solve strongly correlated electronic structure made in \cite{Lee2023Evaluating}) has never been successfully applied to molecular electronic structure due to its prohibitively high scaling.
This brings us to a more pragmatic perspective. For certain high interest systems like FeMoco, we know that despite intense effort, no classical method has solved its electronic structure to the accuracy needed for definitive chemical insight \cite{Berry2025Rapid, Zhai2021Low, Xiang2024Distributed}. We also know that we can efficiently prepare quantum states with a very high overlap on its true ground state \cite{Berry2025Rapid}. Therefore, while the theoretical debate over exponential quantum speedup remains unresolved, having a large-scale quantum computer today would allow us to learn things about the molecule that are currently beyond our reach.

Our team has had success identifying specific, impactful use cases for quantum simulation by collaborating directly with domain experts. This strategy has led to concrete Stage III case studies, including partnerships with pharmaceutical company Boehringer-Ingelheim on drug discovery \cite{Goings2022Reliably}, chemical company BASF on battery development \cite{Rubin2023Fault}, and Sandia National Laboratories on simulating fusion reactors \cite{Rubin2024Quantum}. In the next section, we summarize these and other similar case studies on the resources required for industrial quantum simulation problems in \Cref{tab:stage4_simulation}.

Looking further, we expect that quantum simulation will open new fields of discovery. Our modern world is built on a century of understanding equilibrium quantum mechanics, a domain often tractable by classical methods. In contrast, classical methods struggle to treat non-equilibrium dynamics and quantum computers promise to help open this field by making such simulations routine. This capability may someday allow us to design new phases of matter, like driven superconductors, that only exist far from equilibrium \cite{Fausti2011Light, Mi2022Time}. Similarly, we know that it is possible to program chemical reactions by engineering laser pulses, but simulating such a process is intractable for all but the smallest molecules \cite{Rabitz2000Whither, Zewail2000Femtochemistry}. This is a speculative, long-term vision, and it is difficult to estimate the value of as-yet-undiscovered applications. Nonetheless, we hope that the community will eventually focus as much attention on exploring and realizing such transformative applications as it has on developing techniques suitable for equilibrium systems. Realizing this potential requires the same pragmatic, collaborative work outlined above: e.g., partnering with domain experts to identify the first concrete, impactful simulation problems in this new, unexplored territory.

\section*{Stage IV research}\label{sec:stage_iv}

Once a compelling problem is identified from Stage II or III, the focus shifts to a natural question: what are the minimum physical resources required to achieve quantum advantage? Accurate resource estimates are essential to assessing when quantum computers may deliver a return on investment. Stage IV research answers this by optimizing algorithms and compiling them down to a physical hardware level to produce detailed resource estimates.
This process can be understood as a compilation stack with distinct layers that we outline below.

\begin{figure*}[t]
    \centering
        \includegraphics[width=0.48\textwidth]{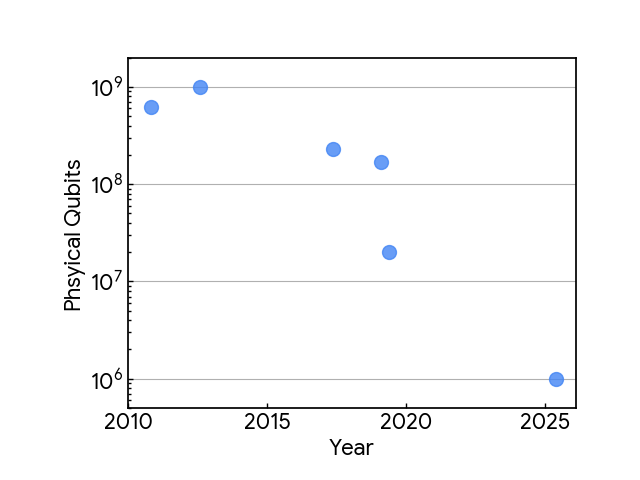}
        \includegraphics[width=0.48\textwidth]{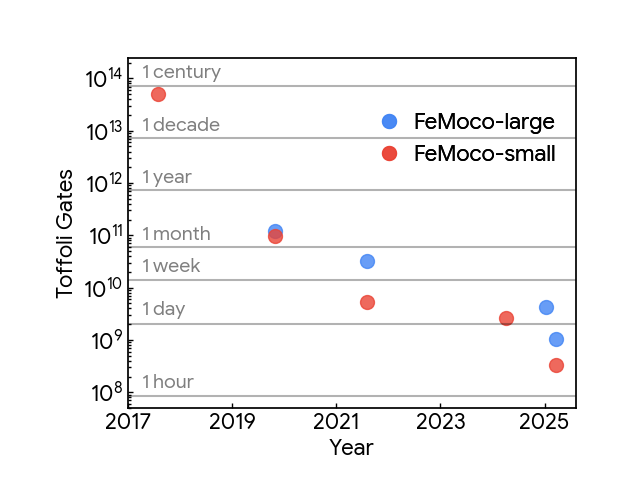}
        \vspace{-0.5cm}
    \caption{These figures illustrate that Stage IV research has reduced the resources required to solve important problems by many orders of magnitude over the last decade as function of publication year in both space such as for breaking 2048 bit RSA encryption (left), and time as approximated by the Toffoli count estimating the ground state energy of FeMoco to chemical accuracy for ``small'' $54$-orbital \cite{Reiher2017Elucidating} and ``large'' $76$-orbital active spaces \cite{Li2019Electronic} (right). The left plot includes  Refs.~\cite{jones2012layered,fowler2012surface,o2017quantum,gheorghiu2019benchmarking,Gidney2025Factoring} and the right plot includes  Refs.~\cite{Reiher2017Elucidating,Lee2020hypercontraction,vonBurg2021Quantum,rocca2024reducing,Low2025Fast}.}   \label{fig:resources}
\end{figure*}
\begin{itemize}
    \item \emph{Stage IV(a): Algorithmic refinement.} The high-level algorithm is refined to better suit the specific use case, often by exploiting symmetries to improve asymptotic performance.
    \item \emph{Stage IV(b): Compilation to logical quantum gates.} The refined algorithm is compiled into a discrete, logical gate set. At this layer, the focus is on constant factors, but may not fully account for the overhead of routing or the production of non-Clifford resources.
    \item \emph{Stage IV(c): Compilation to logical Quantum Error Correction (QEC) primitives.} The gate-level logical circuit is mapped onto a specific architecture, such as the surface code with chip modularity. This produces an instruction set of fault-tolerant operations, such as lattice surgery primitives. At this point one can make physical resource estimates (e.g., number of physical qubits and runtime) by using an appropriate cost model to determine code distances and cycle times.
    \item \emph{Stage IV(d): Compilation to physical operations.} Final machine instructions are optimized and output by an automated compiler. This step produces exact resource counts (up to uncertainties from decoding feedback) as opposed to merely estimates. The computation is now ready to execute given sufficiently capable hardware. 
\end{itemize}

\begin{table*}[t]
\footnotesize
\centering
\begin{tabular}{@{} l l l l l l l l @{}}
\toprule
\wcell{1.5cm}{\textbf{Horizon}} &
\wcell{3.8cm}{\textbf{Problem}} &
\wcell{3.6cm}{\textbf{Stage III applications}} &
\wcell{2.0cm}{\textbf{Problem size}} &
\multicolumn{2}{c}{\textbf{Stage IV(b)}} &
\multicolumn{2}{c}{\textbf{Stage IV(c)}} \\
& & & & \textbf{Logical q.} & \textbf{Toffolis} & \textbf{Phys.~q.} & \textbf{Time} \\
\midrule

\multirow{5}{*}{\wcell{1.5cm}{Medium}} &
\wcell{3.8cm}{Trapdoor claw-free functions~\cite{kahanamoku2022classically}} &
\multirow{2}{*}{\wcell{3.6cm}{Classically verifiable\\quantum advantage}} &
\wcell{2.0cm}{1024 bits} &
1600~\cite{kahanamoku2024fast} & $8\cdot10^{6}$ & -- & -- \\
\cmidrule(lr){2-2}\cmidrule(lr){4-8}
& \wcell{3.8cm}{DQI for OPI~\cite{Jordan2025Optimization}} &
& \wcell{2.0cm}{4095 variables\\70 constraints} &
1900 & $6\cdot10^{6}$ & $8\cdot10^{5}$~\cite{khattar2025verifiable} & 1 hour \\
\cmidrule(lr){2-8}
& \wcell{3.8cm}{Topological data analysis of random graphs~\cite{Lloyd2016Quantum}} &
\wcell{3.6cm}{Unknown} &
\wcell{2.0cm}{256 vertices\\16-cliques} &
500~\cite{Berry2024Analyzing} & $7\cdot10^{9}$ & -- & -- \\
\cmidrule(lr){2-8}
& \wcell{3.8cm}{Factoring~\cite{Shor1999Polynomial}} &
\wcell{3.6cm}{Breaking RSA} &
\wcell{2.0cm}{2048 bits} &
1400 & $7\cdot10^{9}$ & $1\cdot10^{6}$~\cite{Gidney2025Factoring} & 1 week \\
\cmidrule(lr){2-8}
& \wcell{3.8cm}{Binary discrete log~\cite{Shor1999Polynomial}} &
\wcell{3.6cm}{Breaking outdated cryptosystems} &
\wcell{2.0cm}{233 bits} &
3000 & $4\cdot10^{6}$ & $4\cdot10^{6}$~\cite{Garn2025Quantum} & 8 min \\
\midrule

\multirow{4}{*}{\wcell{1.5cm}{Furthest}} &
\wcell{3.8cm}{Elliptic curve cryptography~\cite{Shor1999Polynomial}} &
\wcell{3.6cm}{Breaking elliptic curve cryptography} &
\wcell{2.0cm}{256 bits} &
3000 & $5\cdot10^{7}$ & $9\cdot10^{6}$~\cite{Litinski2023How} & 4 hours \\
\cmidrule(lr){2-8}
& \wcell{3.8cm}{Tensor PCA~\cite{Hastings2020Classical}} &
\wcell{3.6cm}{Unknown} &
\multirow{2}{*}{\wcell{2.0cm}{100 variables}} &
\multirow{2}{*}{900~\cite{fontana2025end}} &
\multirow{2}{*}{$1\cdot10^{15}$} &
\multirow{2}{*}{--} &
\multirow{2}{*}{--} \\
\cmidrule(lr){2-3}
& \wcell{3.8cm}{XOR-SAT refutation~\cite{Schmidhuber2025Quartic}} &
\wcell{3.6cm}{Unknown} &
& & & & \\
\cmidrule(lr){2-8}
& \wcell{3.8cm}{QAOA for max 8-SAT~\cite{Boulebnane2024Solving}} &
\wcell{3.6cm}{Unknown} &
\wcell{2.0cm}{179 variables} &
28000 & $5\cdot10^{11}$ & $7\cdot10^{7}$~\cite{omanakuttan2025threshold} & 15 hours \\
\bottomrule
\end{tabular}
\caption{Algorithms that have reached Stage IV outside of quantum simulation. We summarize just one of the resource estimates from each paper. When physical resource estimates are given for multiple architectures, we focus on reporting numbers for superconducting qubits running the surface code. Still, details of the resource estimates may differ in a way that makes direct comparison challenging. For example, some resource estimates might focus on parallelization to decrease runtime at the cost of additional space. Some detailed resource estimates not included in this table because they involve only quadratic speedup are Refs.~\cite{chakrabarti2021threshold,sanders2020compilation,penuel2024feasibility,dalzell2023end}.}
\label{tab:stage4_not_simulation}
\end{table*}

Stage IV research is most active for the field's mature applications: quantum chemistry and cryptanalysis. Progress in these areas is evident in \Cref{fig:resources} showing the steady, year-over-year reduction in their estimated resource costs, driven by a cycle of algorithmic and compilation improvements that may dramatically shorten the expected time horizon of economic impact. 
Stage IV work sometimes commences with reasonable motivation before a real-world application is found (Stage III). This is especially true for algorithms that could serve as algorithmic tests of quantumness, or hardware demonstrations. However, it is premature to begin resource estimation before establishing an advantage in the first place (Stage II) since interesting resource estimations should be based on instances with quantum advantage.

This work often begins at Stage IV(a), which focuses on refining an algorithm's asymptotic scaling. Such improvements can come from integrating new techniques, like adapting block encodings to chemistry phase estimation \cite{Berry2019Qubitization}. Another common example of Stage IV(a) work involves adapting algorithms to more efficient representations for a target problem, such as employing a residue number system for factoring \cite{Chevignard2025Reducing} or a first-quantized basis in chemistry \cite{Babbush2018Low, Motta2021Low, Babbush2019Quantum}.

To optimize resources required for an application, one must use cost models that become progressively more specific and realistic as research moves through the substages of Stage IV. In Stage IV(b), high-level metrics like ``gate count'' or ``gate depth'' may be acceptable. A standard, more realistic step has been compiling to a minimum number of discrete logical gate set like Clifford + T, though recent advances have called into question the primacy of this cost model, diminishing the historical cost gap between T gates and Clifford gates \cite{Gidney2024Magic}.

As work advances to Stage IV(c), cost models must sensitively depend on the specific error-correction architecture and overhead associated with a potential modularity, with the goal of compiling to abstractions closer to physical implementation such as lattice surgery. Final physical qubit and runtime estimates are then obtained from high-level error-correction parameters like code distances and cycle times. These are derived from extensive numerical simulation of a physical-level error model specific to the hardware platform, and are typically out of the algorithmists' scope. Finally, in Stage IV(d), one uses automatic tools such as compilers to output a complete set of machine instructions for hardware execution.

Making the most effective use of quantum resources requires each stage to be performed with some understanding of cost models for later stages.
For instance, it is common to compile Stage IV(b) to logical quantum gates with all-to-all connectivity, and apply generic qubit routing techniques or other constructive methods like the ``game of surface codes'' to obtain a Stage IV(c) compilation \cite{Litinski2018Game}. 
However, these provide resource upper bounds that are potentially quite loose.
In contrast, we have found better success in approaching Stage IV(b) with mindfulness of gate-connectivity constraints and careful selection of non-Clifford resources~\cite{Lee2020hypercontraction,Gidney2025Factoring}.
In this regard, a deeper understanding of the compilation stack greatly benefits algorithmists seeking resource estimates representative of true hardware capabilities, especially when resources like dense quantum memories \cite{Gidney2025Yoked} and limited non-locality are available.

Currently, most Stage IV resource estimates are performed using a manual process for Stages IV(b) and IV(c). This can be extraordinarily tedious, difficult to verify, and imprecise. As a case in point, several researchers from the authors' team worked for well over a year on an 80-page paper \cite{Su2021Fault} just to determine the constant factors in an algorithm that was introduced and described asymptotically in a preceding 7-page paper \cite{Babbush2019Quantum}.
Due to the time-consuming and architecture-dependent nature of obtaining an optimized Stage IV(d) compilation, it is common for many resource estimates to stop at Stage IV(b).
Keeping in mind that Stage IV is also a moving target that often lags behind the most current techniques, we present selected case studies for algorithms outside quantum simulation in \Cref{tab:stage4_not_simulation}
and inside quantum simulation in \Cref{tab:stage4_simulation}. 
 
 A vital part of Stage IV, adjacent to these substages, is further development of software tools and methods to automate and facilitate compilation and architecture-aware resource estimates.
Current examples include Qualtran \cite{Harrigan2024Expressing}, Q\# \cite{Singhal2022Q}, QREF \cite{PsiQ_QREF}, Bartiq \cite{PsiQ_Bartiq}, Qrisp \cite{Seidel2024Qrisp}, and Silq \cite{Bichsel2020Silq} to compose and compile quantum algorithms, with automatic translation of a logical algorithm directly into a full physical instruction set. This will not only provide precise resource counts but will also enable a workflow analogous to modern software profiling. Researchers will be able to test different implementations, identify performance bottlenecks, empirically optimize their quantum code, and verify their compilations.

\begin{table*}[!t]
\scriptsize
\centering
\setlength{\tabcolsep}{3.6pt}
\renewcommand{\arraystretch}{1.12}
\begin{tabular}{@{} l l l l l l l l l @{}}
\toprule
\wcell{1.1cm}{\textbf{Horizon}} &
\wcell{2.8cm}{\textbf{Model\\problem}} &
\wcell{3.0cm}{\textbf{Stage III\\applications}} &
\wcell{2.0cm}{\textbf{Problem\\size}} &
\wcell{2.0cm}{\textbf{Quantum advantage}} &
\multicolumn{2}{c}{\textbf{Stage IV(b)}} &
\multicolumn{2}{c}{\textbf{Stage IV(c)}} \\
& & & & & \wcell{1.5cm}{\textbf{Logical q.}} & \textbf{Toffolis} & \textbf{Phys.~q.} & \textbf{Time} \\
\midrule

\multirow{5}{*}{\wcell{1.1cm}{Nearest}} &
\multirow{7}{*}{\wcell{2.8cm}{Condensed matter physics models}} &
\wcell{3.0cm}{2D Ising quench\textdagger} &
\wcell{2.0cm}{121 sites} &
\multirow{2}{*}{\wcell{2.0cm}{Evaluated vs.\ MPS~\cite{Mandra2025}}} &
140~\cite{Huggins2025} & $4\cdot 10^{4}$ & $6\cdot 10^{4}$ & 1 second \\
\cmidrule(lr){3-4}\cmidrule(lr){6-9}
&
&
\wcell{3.0cm}{2D Hubbard quench\textdagger} &
\wcell{2.0cm}{128 spin orbitals} &
&
160~\cite{kan2025resource} & $1\cdot 10^{6}$ & -- & -- \\
\cmidrule(lr){3-9}
&
&
\wcell{3.0cm}{Heisenberg quench\textdagger} &
\wcell{2.0cm}{70 sites} &
\wcell{2.0cm}{Evaluated vs.\ exact sim.} &
131~\cite{nam2019low} & $6\cdot 10^{6}$ & -- & -- \\
\cmidrule(lr){3-9}
&
&
\wcell{3.0cm}{3D uniform electron\\gas ground state} &
\wcell{2.0cm}{64 electrons\\64 orbitals} &
\multirow{2}{*}{\wcell{2.0cm}{Evaluated vs.\ QMC~\cite{babbush2018encoding}}} &
132~\cite{kivlichan2020improved} & $4\cdot 10^{7}$ & $3\cdot 10^{5}$ & 4 hours \\
\cmidrule(lr){3-4}\cmidrule(lr){6-9}
&
&
\wcell{3.0cm}{2D uniform electron\\gas ground state} &
\wcell{2.0cm}{49 electrons\\64 orbitals} &
&
140~\cite{mcardle2022exploiting} & $7\cdot 10^{8}$ & -- & -- \\
\cmidrule(lr){1-1}\cmidrule(lr){3-9}

\multirow{8}{*}{\wcell{1.1cm}{Medium}} &
&
\wcell{3.0cm}{NMR spectroscopy\textdagger} &
\wcell{2.0cm}{32 spins} &
\wcell{2.0cm}{Plausible vs.\ TEBD~\cite{elenewski2024prospects}} &
100~\cite{elenewski2024prospects} & $3\cdot 10^{7}$ & -- & -- \\
\cmidrule(lr){3-9}
&
&
\wcell{3.0cm}{Sachdev--Kitaev--Ye\textdagger} &
\wcell{2.0cm}{200 sites} &
\wcell{2.0cm}{Plausible but not assessed} &
300~\cite{babbush2019quantum2} & $3\cdot 10^{7}$ & -- & -- \\
\cmidrule(lr){2-9}
& \wcell{2.8cm}{Chemical dynamics\\(2nd quantization + vibrations)} &
\wcell{3.0cm}{Singlet-fission organic\\solar cells\textdagger~\cite{motlagh2025quantum}} &
\wcell{2.0cm}{6 states\\21 modes} &
\wcell{2.0cm}{Evaluated vs.\ MCTDH~\cite{xie2015full}} &
154~\cite{motlagh2025quantum} & $3\cdot 10^{6}$ & -- & -- \\
\cmidrule(lr){2-9}
& \wcell{2.8cm}{Nuclear dynamics\\(1st quantization)} &
\wcell{3.0cm}{Pionless effective\\field theory\textdagger} &
\wcell{2.0cm}{40 nucleons} &
\wcell{2.0cm}{Plausible but not assessed} &
500~\cite{spagnoli2025quantum} & $6\cdot 10^{8}$ & -- & -- \\
\cmidrule(lr){2-9}
& \wcell{2.8cm}{X-ray absorption\\
spectroscopy\\(2nd quantization)} &
\wcell{3.0cm}{Structure of Li$_2$MnO$_3$ battery cathode~\cite{kunitsa2025quantum}} &
\wcell{2.0cm}{18 orbitals} &
\wcell{2.0cm}{Plausible beyond RAS methods~\cite{fomichev2025fast}} &
100~\cite{fomichev2025fast} & $5\cdot 10^{11}$ & -- & -- \\
\cmidrule(lr){2-9}

& \multirow{5}{*}{\wcell{2.8cm}{Chemical\\ ground states\\(2nd quantization)}} &
\wcell{3.0cm}{Carbon fixation by\\Ru-complex~\cite{vonBurg2021Quantum}} &
\wcell{2.0cm}{56 orbitals} &
\wcell{2.0cm}{Not assessed} &
900~\cite{Low2025Fast} & $2\cdot 10^{8}$ & -- & -- \\
\cmidrule(lr){3-9}
&
&
\wcell{3.0cm}{Oxidation by p450 enzyme~\cite{Goings2022Reliably}} &
\wcell{2.0cm}{58 orbitals} &
\wcell{2.0cm}{Evaluated vs.\ DMRG~\cite{Goings2022Reliably}} &
1200~\cite{Low2025Fast} & $5\cdot 10^{8}$ & -- & -- \\
\cmidrule(lr){3-9}
&
&
\wcell{3.0cm}{Nitrogen fixation by\\FeMo-cofactor~\cite{Reiher2017Elucidating}} &
\wcell{2.0cm}{76 orbitals} &
\wcell{2.0cm}{Evaluated vs.\ DMRG~\cite{Berry2025Rapid}} &
1500~\cite{Low2025Fast} & $1\cdot 10^{9}$ & $5\cdot 10^{6}$ & 9 hours \\
\cmidrule(lr){1-1}\cmidrule(lr){3-9}

\multirow{7}{*}{\wcell{1.1cm}{Furthest}} &
&
\wcell{3.0cm}{Structure of LiPF$_6$\\battery electrolyte~\cite{kim2022fault}} &
\wcell{2.0cm}{116 orbitals} &
\wcell{2.0cm}{Plausible but not assessed} &
18000~\cite{kim2022fault} & $9\cdot 10^{11}$ & -- & -- \\
\cmidrule(lr){3-9}
&
&
\wcell{3.0cm}{Bridged $\textrm{Mo}_2$~\cite{bellonzi2024feasibility}} &
\wcell{2.0cm}{70 orbitals} &
\wcell{2.0cm}{Evaluated vs.\ DMRG~\cite{bellonzi2024feasibility}} &
4100~\cite{bellonzi2024feasibility} & $3\cdot 10^{12}$ & $6\cdot 10^{7}$ & 2 years \\
\cmidrule(lr){2-9}
& \multirow{2}{*}{\wcell{2.8cm}{Chemical dynamics\\(1st quantization)}} &
\wcell{3.0cm}{Reaction of NH$_3$\\ and BF$_3$~\cite{da2025comprehensive}} &
\wcell{2.0cm}{40 electrons} &
\wcell{2.0cm}{Plausible but not assessed} &
\wcell{1.55cm}{800~\cite{da2025comprehensive} w.o.~ancillae} & $2\cdot 10^{15}$ & -- & -- \\
\cmidrule(lr){3-9}
&
&
\wcell{3.0cm}{Stopping power for\\inertial fusion~\cite{Rubin2024Quantum}} &
\wcell{2.0cm}{218 electrons} &
\wcell{2.0cm}{Plausible beyond TDDFT} &
6200~\cite{Rubin2024Quantum} & $1\cdot 10^{15}$ & -- & -- \\
\cmidrule(lr){2-9}
& \wcell{2.8cm}{Material ground states\\(1st quantization)} &
\multirow{2}{*}{\wcell{3.0cm}{Structure of LiNiO$_2$\\battery cathode~\cite{Rubin2023Fault}}} &
\multirow{2}{*}{\wcell{2.0cm}{92 electrons\\58 orbitals\\$[2,2,1]$ supercell}} &
\multirow{2}{*}{\wcell{2.0cm}{Plausible beyond CCSD}} &
1400~\cite{Berry2024Quantum} & $1\cdot 10^{14}$ & -- & -- \\
\cmidrule(lr){2-2}\cmidrule(lr){6-9}
& \multirow{2}{*}{\wcell{2.8cm}{Material ground states\\(2nd quantization)}} &
&
&
&
80000~\cite{Rubin2023Fault} & $1\cdot 10^{12}$ & $9\cdot 10^{7}$ & 9 months \\
\cmidrule(lr){3-9}
&
&
\wcell{3.0cm}{Heterogeneous NiO / PdO catalysis~\cite{ivanov2023quantum}} &
\wcell{2.0cm}{64–72 atom supercell} &
\wcell{2.0cm}{Plausible but not assessed} &
$100000$~\cite{ivanov2023quantum} & $3\cdot 10^{11}$ & $2\cdot 10^{8}$ & 3 months \\
\bottomrule
\end{tabular}
\caption{\small Non-comprehensive compendium of Stage IV research progress in quantum simulation. Dynamics simulation estimates include number of repetitions required to estimate an observable unless marked by \textdagger\, which indicates that estimates are only for a single shot. Applications are roughly classified into different time horizons based on the reported number of physical $10^{-3}$ error-rate superconducting qubits (nearest: $\sim 10^5$, medium: $\sim 10^6$, furthest: $\geq 10^7$) needed for a first demonstration of useful quantum advantage on a surface code. One thing to note is that while there are many high value applications requiring the simulation of \emph{ab initio} solid-state models, such simulations are quite expensive and more research is needed to bring down costs. Where unavailable, we compare across different architectures by making a subjective classification based on the reported number of logical qubits and Toffoli gates (or equivalent), which may not capture routing overheads, opportunities to parallelize non-Clifford operations, or other space-time tradeoffs. Some other resource estimates worth mentioning include Refs.~\cite{agrawal2024quantifying,nguyen2024quantum,watts2024fullerene,clinton2024towards,malpathak2025trotter,baker2024spindefect,loaiza202infraredspectroscopy,casares2025quantumalgorithmsdetectodmractive}.}
\label{tab:stage4_simulation}
\end{table*}

\subsection*{Compilation for early fault-tolerance}

As quantum hardware scales, the field is rapidly approaching the era of early fault-tolerance when quantum error correction first begins to enable larger-scale quantum computation. This transition raises a crucial question: What will be the first compelling applications to run on these machines? Several key milestones will define this new era, including, e.g.:
\begin{itemize}
    \item The first QEC demonstration solving a classically intractable non-verifiable problem.
    \item The first QEC demonstration that is classically intractable and quantumly verifiable.
    \item The first scientific discovery enabled using fault tolerance for quantum simulation.
    \item The first QEC demonstration of an algorithm that is classically verifiable.
\end{itemize}

There is significant confusion in the literature and little consensus on how to evaluate algorithms for this era of early fault-tolerance. Even a seemingly simple question---such as whether algorithmic technique X is better than technique Y for this era---can be divisive. For example, it is often suggested that one should aim for fewer logical qubits \cite{Lin2022Heisenberg, Katabarwa2024Early}. However, if this increases the number of gates, then the code distance may need to increase, which could result in a larger overall physical qubit requirement. While the primary goal is usually to minimize the required number of physical qubits, achieving this can be complex. Final resource counts depend sensitively on the chosen QEC architecture, and the detailed cost models needed to truly differentiate approaches. This challenge is compounded by the fact that the underlying platform is a moving target: ideas about error-correcting codes, hardware architectures, and compilation tools are all still evolving at a rapid pace. In the era of early fault-tolerance, it will be critical to co-design quantum algorithms and the specifics of an error-correction architecture. Even hardware decisions such as how to best modularize chips can ultimately play an important role in what sort of tradeoffs are advantageous.

As an example, consider optimizing for circuit depth. Depth only translates to time if operations scheduled simultaneously in a logical layer can actually execute concurrently. However, one quickly runs into hard limitations in the practical ability to route resource states, and to decode and implement feedback correction, as required for non-Clifford gates in most error-correcting codes. In the regime of early fault-tolerance, it is more likely that one is distilling a smaller number of resource states in the first place in order to save physical qubits, which would require non-Clifford gate serialization. Even when layers consist only of Clifford gates, these gates typically require non-trivial ancilla space to execute (e.g., in the standard surface code), and so parallelizing operations (as one is apt to do when aiming for lower circuit depth) may increase the number of ancillas required and thus, the total number of physical qubits.

As another example, consider estimating an expectation value to precision $\epsilon$ for a circuit with $G$ gates. Amplitude estimation requires executing a single deep circuit with ${\cal O}(G/\epsilon)$ gates, whereas sampling requires ${\cal O}(1/\epsilon^2)$ repetitions of the shallow ($G$ gates) circuit. Intuition might suggest that the shallow circuits of the sampling approach are better for early fault-tolerant hardware, but this intuition is incomplete. Amplitude estimation requires a logical gate error rate of ${\cal O}(\epsilon/G)$ to succeed. A naive sampling approach requires a similar error rate. At this error rate, the probability of a logical gate error in any individual sample is $\sim\epsilon$. This introduces a bias in the estimated expectation value on the order of $\epsilon$, and any higher error rate would introduce an unacceptably larger bias. Therefore, naive sampling demands the same code distance as amplitude estimation and a higher total runtime. While error mitigation can relax this requirement, added circuit repetitions required for error mitigation may favor amplitude estimation for achieving practical runtimes. Ultimately, a detailed study is required to assess which approach is more favorable, and the conclusions will depend heavily on the specific architecture, target fidelities, and fault-tolerant error models.

A fine-grained understanding of costs in early fault-tolerance will likely require full compilation to a specific architecture. We expect that building out precise compilation stacks will be a major focus of industrial efforts over the next few years. However, high-quality, open-source compilers will likely remain scarce for academic use, since developing these tools requires a massive software engineering effort, and compilers are necessarily tailored to their target architectures. For example, the underlying connectivity of the hardware can influence both the choice of error correcting code (e.g, the surface code vs an LDPC code) \cite{Bravyi2024High, Bluvstein2022Quantum} and the relative difficulty of implementing certain operations (e.g., increased connectivity can allow for more transversal gates) \cite{Bluvstein2022Quantum, Litinski2022Active}. Photonic architectures offer another example of an interesting architecture-specific capability that any serious compilation effort would take advantage of: the ability to effectively “store” a large number of qubits using long fiber optic delay lines without significantly increasing the cost of the hardware \cite{Bartolucci2023Fusion}.

Despite the practical challenges of evaluating and optimizing algorithms for early fault-tolerance, we believe that this is an area of research rich with opportunities. For example, compiling to abstractions like braiding diagrams, lattice surgery, or modified ZX calculus offers a valuable middle ground between logical and physical circuits. Other efforts may make progress by focusing on specific architectural choices. Our team has constructed a rough cost model for early fault-tolerance that we call FLASQ (fluid allocation of surface code qubits) \cite{Huggins2025}. Designed to produce an estimate of surface code resources given logical circuits compiled to a 2D grid, we hope it will be useful for quantum algorithms researchers not willing to engage as deeply with the details of quantum error-correction. However, we conclude by noting that algorithmists who do embrace the messiness of fault-tolerance have a rare opportunity in the coming years to play an important role in shaping hardware through co-design by conducting studies that reveal the impact of hardware architecture (e.g., how chips are modularized) on algorithmic overhead.

\section*{The economics of quantum application discovery}

The promise of quantum computing is tempered by the immaturity of both its hardware and applications. At this nascent stage, continued progress depends on sustained investment from public and private sources. For these stakeholders, who weigh opportunities based on risk and reward, the challenge of building scalable, fault-tolerant hardware is the primary source of risk. Consequently, tangible hardware progress (achieving higher qubit counts, lower error rates, scalable error correction experiments, and other critical milestones) helps derisk the endeavor. Such derisking solidifies confidence that large-scale, error-corrected quantum computers are an achievable goal, reassuring investors and securing the long-term vision of the field.

On the other side of the investment equation is the ``reward'': the commercial or scientific value and revenue opportunities driven by applications with quantum advantage. The projected size of this market (e.g., whether billions or hundreds of billions of dollars annually) directly influences company valuations and strengthens the case for investment. Investors may currently operate under the assumption that ``if we build it, apps will come.'' For that narrative to remain compelling, however, the rate of application discovery must accelerate as hardware matures. A perceived slowdown may be interpreted as a signal that the technology's future value is limited to our current ensemble of established quantum use cases, diminishing the expected reward.

For any stakeholder---be they a government agency, venture capitalist, or corporate executive---the ``reward'' in quantum's risk/reward ratio is best articulated through specific and credible application case studies. An ideal case study presents a broad use case, grounds it in at least one specific example with quantifiable value, and assesses the quantum resources required for quantum advantage. At present, the few such comprehensive case studies that have been published are largely confined to quantum simulation and cryptanalysis (we summarize much of this work in the tables of the preceding section).

These days, many papers on the quant-ph arXiv explore the topic of quantum applications by relating established quantum algorithms to industrial challenges while avoiding the singular question that relentlessly haunts our collective endeavor: is this likely to ever lead to a quantum advantage? For the most part, hype in the quantum applications space is easy for practicing researchers to identify and ignore. A bigger problem is that the noise created by such work can lead to confusion in policy and business circles. Researchers sometimes imagine that corporate executives and venture capitalists uncritically believe everything they hear about quantum computing. In reality, deep tech investors are extremely familiar with hype around new technologies. Nevertheless, our field may burn through valuable public trust if exaggerated narratives around timelines and applications of quantum computing perpetuate unchecked.

While it is important for scientists to push back on hype in the quantum field, we also caution against overcompensating and effectively ``hyping in the opposite direction'' (i.e., spreading misleading information, or employing hyperbole in service of combating hype). Such a strategy can lead to excessively cynical takes coming from parts of the academic community, discouraging students from working on quantum applications or besmirching researchers trying to push forward an optimistic yet scientifically grounded research vision. We need top researchers working on quantum applications and attitudes that treat all work on quantum applications as inherently suspect are detrimental towards that end.

Finally, when it comes to the hard work required to uncover and substantiate truly promising quantum applications, we believe the quantum industry appears to face a classic collective action problem, leading to systemic under-investment in high quality work in this area. While the necessity of new applications is widely acknowledged, prevailing wisdom is that the competitive advantage of major quantum efforts lies primarily in hardware. This view is often paired with a belief that intellectual property for algorithms is either marginal or could be relatively easily circumvented. The resulting incentive structure creates an unfortunate Nash Equilibrium: the dominant strategy for any individual company is to focus an outsized share of resources on a defensible hardware moat while assuming other groups, or academics, shoulder the burden of serious application discovery.

\section*{Outlook}

The journey from a theoretical quantum algorithm to a practical, deployed application is a long and complex endeavor. In this perspective, we have introduced a five-stage framework to categorize and understand the maturity of applications research, arguing that progress is not uniform across all stages. While the community has seen success in the discovery of abstract algorithms (Stage I) and the detailed optimization of a few select use cases (Stage IV), the field's forward momentum is limited by underappreciated bottlenecks in the middle stages. Specifically, the challenges of identifying concrete problem instances with quantum advantage (Stage II) and matching these problems to real-world applications (Stage III) represent the most significant hurdles to expanding the landscape of quantum applications.

Successfully navigating these stages requires confronting both technical and sociological challenges. The technical difficulty of Stage II often lies in the delicate balance of finding problem instances that cause the quantum algorithm to transition through states complex enough to be classically intractable, yet structured enough to prevent the quantum computer's output from concentrating to trivial, classically predictable values. The primary challenge of Stage III is due in part to the narrow window of quantum advantage that survives the first two stages and is in part sociological: a knowledge gap exists between the experts who understand the nuanced requirements for quantum advantage and the domain specialists who know where such problem structures might appear in the real world.

We believe, however, that these challenges are surmountable with a strategic and collaborative effort. Ultimately, the continued justification for the  investment required to build fault-tolerant quantum computers depends on our ability to articulate a growing portfolio of valuable, achievable, and specific applications. By harnessing the community's ingenuity, we can accelerate the discovery of these applications and bring the transformative potential of quantum computing into clearer view.

\subsection*{Acknowledgements}

The authors thank the following people for helpful discussions or feedback which helped shape our perspective on these matters: Garnet Chan, David Gosset, Matt Harrigan, Sid Jain, Stephen Jordan, Robin Kothari, Tony Metger, Danial Motlagh, Hartmut Neven, Matt Reagor, 
Juan Miguel Arrazola, Alexander Schmidhuber, Vadim Smelyanskiy, Rolando Somma, Guifre Vidal, Alec White, Adam Zalcman and Mark Zhandry.

\raggedbottom
\pagebreak
\bibliography{refs}

\end{document}